# A New Algorithm for Pyramidal Clipping of Line Segments in $E^3$

Vaclav Skala[1], Duc Huy Bui
Department of Informatics and Computer Science[2]
University of West Bohemia
Univerzitni 22, Box 314
306 14 Plzen
Czech Republic

{skala | bui} @kiv.zcu.cz        http://iason.zcu.cz/{~skala | ~bui}

**Abstract**

A new algorithm for clipping a line segment against a pyramid in $E^3$ is presented. This algorithm avoids computation of intersection points which are not end-points of the output line segment. It also allows solving all cases more effectively. The performance of this algorithm is shown to be consistently better than existing algorithms, including the Cohen-Sutherland, Liang-Barsky and Cyrus-Beck algorithms.

## 1. Introduction

Let us assume that two points $A(x_A, y_A, z_A)$ and $B(x_B, y_B, z_B)$ are given and we wish to compute the intersection of the line segment $AB$ with the unitary clipping pyramid, defined as the set of all points $(x, y, z)$ such that $-z \leq x \leq z$ and $-z \leq y \leq z$ $(z \geq 0)$. The intersection is either empty or a line segment whose end-points we must compute.

Many algorithms for clipping a line or a line segment in $E^3$ have been published, see standard textbooks (Cyrus and Beck 1978; Liang and Barsky 1983-1984; Foley, van Dam, Feiner and Hughes 1990; Skala 1996; Skala 1997) for main references. For a long time the Cohen-Sutherland algorithm (CS) and its extensions to $E^3$ (Foley, van Dam, Feiner and Hughes 1990) were the only line segment clipping algorithms found in most textbooks. Later, the Liang-Barsky (LB) and Cyrus-Beck (CB) algorithms were proposed and are based on the line parametric representation.

Before describing the proposed algorithms for line segment clipping, it is necessary to unify some definitions.

## 2. Definitions

The planes $x = -z$, $x = z$, $y = -z$ and $y = z$ are called the right, left, bottom and top boundaries of the unitary pyramid, respectively. We will say that:
- a point or a line segment is visible, if it lies entirely inside the given pyramid,
- a point or a line segment is invisible, if it lies entirely outside the given pyramid,
- a line segment is partially visible, if it lies partly inside the given pyramid and partly outside.

If a line segment is invisible, then no part of the line segment appears in the output, and the line segment is said to be rejected by the clipping algorithm.

The boundaries of the pyramid divide the Cartesian positive half-space ($z \geq 0$) into 9 regions. Regions which are bounded by only two boundaries are called the corner regions and regions which are bounded by three boundaries are called the edge regions, see Figure 1.

## 3. Proposed Pyramidal clipping algorithm

---

[1] Affiliated with the Multimedia Technology Research Centre, University of Bath, U.K.
[2] This work was supported by The Ministry of Education of the Czech Republic: project VS 97155 and project GA AV A2030801.

The proposed Pyramidal clipping algorithm (PC) uses a similar approach as Nicholl-Lee-Nicholl algorithm (Nicholl, Lee and Nicholl 1987) that was derived for $E^2$ case only. Given a line segment AB, for simplicity we assume that both points A and B are in the positive half-space. The end-point A therefore can lie inside of the pyramid, in an edge region or in a corner region. For each of these cases, we can divide the positive half-space into certain number of sub-regions, see Figures 2-4. These sub-regions are bounded by the pyramid's boundaries and planes determined by point A and one edge of pyramid. These planes will be denoted $\rho_1$, $\rho_2$, $\rho_3$, $\rho_4$, clockwise from the top-left edge, see Figures 2-4. All other cases can be obtained from one of these cases in Figures 2-4 by rotating the scene around the z axis.

With the above definitions the algorithm can be described by the following basic steps:
- characterize the location of point A among the 9 regions,
- characterize the location of point B among the appropriate sub-regions,
- compute the intersection points according to above characterization.

The main advantage of this approach is that we can determine which pyramid boundaries are intersected and therefore avoid unnecessary computation of invalid intersection points.

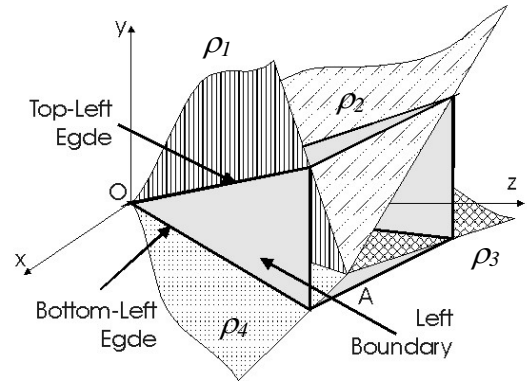

Figure 2: Sub-regions for the case when point A lies inside the pyramid.

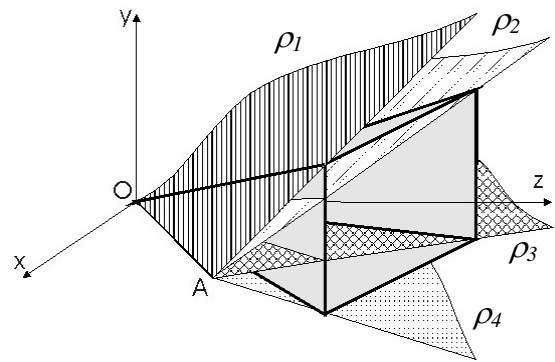

Figure 3: Sub-regions for the case when point A is in left edge region.

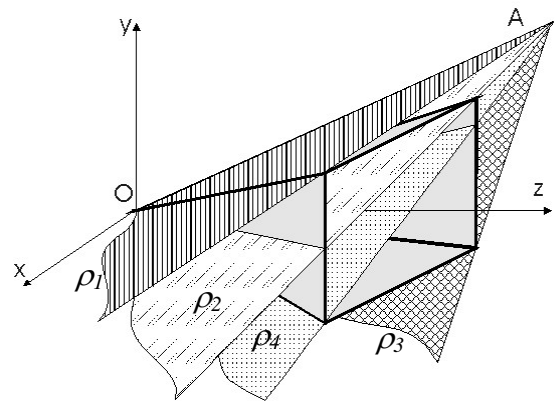

Figure 4: Sub-regions for the case when point A is in top-right corner region.

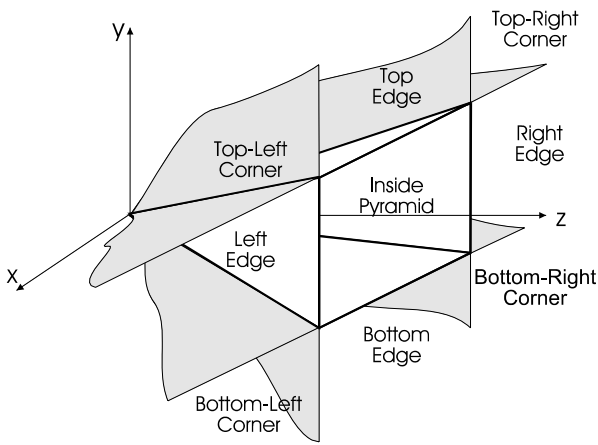

Figure 1: Subdivision of the positive haft-space into regions.

The proposed PC algorithm will be explained in more detail by the top-down approach. At the beginning we must determine whether the end-point A of the given line segment is beyond the right boundary, beyond the left boundary or between those two boundaries. The main procedure, in Pascal-like code is:

**procedure** Clip( $x_A$, $y_A$, $z_A$, $x_B$, $y_B$, $z_B$: **real**);
**begin**
  **if** $x_A$ < -$z_A$ **then**    case_I
        {point A is beyond right boundary}
  **else if** $x_A$ <= $z_A$ **then** case_II
        {point A is between 2 boundaries}
    **else**        case_III
        {point A is beyond left boundary}
**end**;

We will use the EXIT command in the following parts of the algorithm to denote the end of the procedure and to avoid "**else if**" sequences unnecessary for the algorithm explanation.

### I. Point A is beyond the right boundary.

If point B is also beyond the right boundary, it is not necessary to further characterize point A, because the line segment is invisible. Therefore, we must check whether point B is beyond the right boundary before proceeding on. After that, we must test whether point A lies either in the corner region or in the edge region. This section of algorithm can be implemented as follows:

**begin** {case_I}
  **if** $x_B$ < -$z_B$ **then**
        **EXIT**; {Line segment is rejected}
  **if** $y_A$ > $z_A$ **then** case_I_1
        {point A is in the top-right corner region}
  **else if** $y_A$ >= -$z_A$ **then** case_I_2
        {point A is in the right edge region}
    **else** case_I_3
        {point A is in the bottom-right corner region}
**end** {case_I};

### I.1. Point A is in the top-right corner region and point B is not beyond the right boundary.

If point B is above the top boundary, the line segment is invisible and no further computation is needed. Therefore, we need to check this condition first, and then characterize point B so that we can distinguish between the case when point B is beyond the left boundary and the case when point B is inside of the pyramid or in the bottom edge region, see Figure 4. The following pseudo-code shows how it can be implemented:

**begin** {case_I_1}
  **if** $y_B$ > $z_B$ **then**
        **EXIT**; {Line segment is rejected}
  **if** $x_B$ > $z_B$ **then** case_I_1_a
        {point B is in the left edge or in the
          bottom-left corner region}
    **else** case_I_1_b
        {point B is inside of the pyramid or
          in the bottom edge region}
**end** {case_I_1};

### I.1.a) Point A is in the top-right corner region and point B is in the left edge region or in the bottom-left corner region.

If point B is above the plane $\rho_1$, the line segment is rejected, see Figure 4. Therefore, we must check this condition first, and then distinguish the case when point B is in the left edge region and the case when point B is in the bottom-left corner region. In the case of the left edge region, one intersection point lies on the pyramid's left boundary. The location of point B against the plane $\rho_2$ will specify, that the second intersection point lies on the top or on the right boundary. By this way only the appropriate intersection point is computed. In the case of bottom-left corner region, the comparison of point B against the plane $\rho_3$ is performed first to eliminate the case when the line segment is rejected. After that, we compare the position of point B against the plane $\rho_4$ to determine location of the first intersection point (on the bottom or on the left boundary). At the end, a similar comparison with the plane $\rho_2$ is performed to determine the second intersection point. An implementation can be as follows:

```
begin {case_I_1_a}
  Δx := xB – xA;  Δy := yB – yA;  Δz := zB – zA;
  if ((xA–zA)*(Δz-Δy) > (yA-zA)*(Δz-Δx))
     then EXIT; {Line segment is rejected}
  {first intersection point computation}
  if yB > -zB then {B is in the left edge region}
     t1 := (xA–zA)/( Δz-Δx)
  else   {B is in the bottom left corner region}
    begin
       if ((xA+zA)*(Δz+Δy)
         >(yA+zA)*(Δz+Δx))
         then EXIT;{Line segment is rejected}
       if ((zA-xA)*(Δy+Δz) > (zA+yA)*(Δz-Δx))
         then {intersection with left boundary}
            t1 := (xA–zA)/(Δz-Δx)
         else{intersection with bottom boundary}
            t1 := -(yA+zA)/(Δz+Δy)
    end;
  {second intersection point computation}
  if ((xA+zA)*(Δz-Δy) > (zA-yA)*(Δz+Δx))
    then     {intersection with top boundary}
         t2 := (yA–zA)/(Δz-Δy)
    else     {intersection with right boundary}
         t2 := -(xA+zA)/(Δz+Δx)
end {case_I_1_a};
```

### I.1.b) *Point A is in the top-right corner region and point B is inside of the pyramid or in the bottom edge region.*

In the case when point *B* is inside of the pyramid, the position of point *B* against the plane $\rho_2$ specify whether the intersection point lies either on the top or on the right boundary. In the case when point *B* is in the bottom edge region, the comparison of point *B* against the plane $\rho_3$ must be performed first to eliminate the situation when the line segment is rejected. If the line segment *AB* is not rejected by the clipping algorithm then one intersection point lies on the bottom boundary. The other intersection point lies on the top or on the right boundary according to the position of point *B* against the plane $\rho_2$. This section can be implemented as follows:

```
begin {case_I_1_b}
  Δx := xB – xA;  Δy := yB – yA;  Δz := zB – zA;
  {first intersection point computation}
  if yB < -zB then  {B in bottom edge region}
     begin
       if
         ((xA+zA)*(Δz+Δy)>(yA+zA)*(Δz+Δx))
         then EXIT;{Line segment is rejected}
       t1 := -(yA+zA)/(Δz+Δy)
     end
  else  { B is inside the pyramid }
     t1 := 1;
  {second intersection point computation}
  if ((xA+zA)*(Δz-Δy) > (zA-yA)*(Δz+Δx))
     then {intersection with top boundary}
       t2 := (yA–zA)/(Δz-Δy)
     else {intersection with right boundary}
       t2 := -(xA+zA)/(Δz+Δx)
end {case_I_1_b};
```

### I.2. *Point A is in the right edge region and point B is not beyond the right boundary.*

We need to distinguish the cases, when point *B* is bellow the bottom boundary (point *B* is in bottom-left corner region or in the bottom edge region), or above the top boundary (point *B* is in top-left corner region or in the top edge region) or between top and bottom boundaries. An implementation can be as follows:

```
begin {case_I_2}
  if yB <  -zB then case_I_2_a
         {point B is in the bottom-left corner
              or in the bottom edge region}
     else if yB <= zB then case_I_2_b
         {point B is in the left edge region or
                 inside of the pyramid}
         else case_I_2_c
              {point B is in the top-left corner
                    or in the top edge region}
end {case_I_2}
```

### I.2.a) *Point A is in the right edge region and point B is in the bottom-left corner or in the bottom edge region.*

The location of point *B* against the plane $\rho_3$ helps us to eliminate the case when the line segment is rejected. If point *B* is in the bottom edge region then the intersection points are on the right and the bottom boundaries. If point *B* is in the bottom-left corner region then one intersection point lies on the right boundary and the second intersection point's location (either on the left or on the bottom boundary) is determined by the location of point *B* against the plane $\rho_4$.

```
begin {case_I_2_a}
  if ((x_A+z_A)*(Δz+Δy) > (y_A+z_A)*(Δz+Δx))
    then EXIT;{Line segment is rejected}
  {first intersection point computation}
  if x_B > z_B then {B in the bottom-left corner}
    if ((z_A-x_A)*(Δz+Δy) > (z_A+y_A)*(Δz-Δx))
      then {intersection with left boundary}
        t_1:=(x_A–z_A)/(Δz-Δx)
      else{intersection with bottom boundary}
        t_1:= -(y_A+z_A)/(Δz+Δy)
  else {B in bottom edge region}
        {intersection with bottom boundary}
        t_1:= -(y_A+z_A)/(Δz+Δy);
  {second intersection is on right boundary}
  t_2 := -(x_A+z_A)/(Δz+Δx)
end {case_I_2_a};
```

*I.2.b) Point A is in the right edge region and point B is inside of the pyramid or in the left edge region.*

In this case, one intersection point is on the right boundary and the second one (if point *B* is in the left edge region) is on the left boundary, see following pseudo-code:

```
begin {case_I_2_b}
  {first intersection is on right boundary }
  t_1 := -(x_A+z_A)/(Δz+Δx);
  {second intersection point computation}
  if x_B > z_B then {point B in left edge region}
    t_2 := (x_A–z_A)/(Δz-Δx)
  else   t_2:= 1
end {case_I_2_b};
```

*I.2.c) Point A is in the right edge region and point B is in the top-left corner or in the top edge region:* similar to case_I_2_a.

*I.3. Point A is in the bottom-right corner region and point B is not beyond the right boundary.*

The case is similar to case_I_1.

*II. Point A is between the left and the right boundaries.*

In this case, we need to characterize the location of point *A* to specify that, if point *A* lies inside of the pyramid or in an edge region. The following pseudo-code shows how it can be done:

```
begin {case_II}
  if y_A > z_A then case_II_1
              {A is in the top edge region}
  else if y_A < -z_A then case_II_2
              {A is in the bottom edge region}
       else case_II_3{A is inside the pyramid}
end {case_II};
```

We need to consider only the case when point *A* is inside the pyramid (case_II_3). The cases, when point *A* is in the top (case_II_1) or bottom edge region (case_II_2), are similar to the case when point *A* is in the right edge region (case_I_2).

*II.3. Point A is inside the pyramid.*

If point *B* lies in an edge region then the boundary, on which the intersection point lies, is determined, see the Figure 2, and the appropriate intersection point is computed. If point *B* lies in a corner region then a comparison of the location of point *B* with an appropriate plane $\rho_i$ is necessary before the appropriate intersection point is computed. An implementation can be illustrated as follows:

```
begin {case_II_3}
  if x_B < -z_B then
    if y_B > z_B then case_II_3_a
          {B is in the top-right corner region}
    else if y_B >= -z_B then case _II_3_b
          {B is in the right edge region}
       else case_II_3_c
        {B is in the bottom-right corner region}
  else if x_B > z_B then
    if y_B > z_B then case_II_3_d
          {B is in the top-left corner region}
    else if y_B < -z_B then case_II_3_e
        {B is in the bottom-left corner region}
       else case_II_3_f
              {B is in the left edge  region}
  else
    if y_B > z_B then case_II_3_g
```

          {$B$ is in the top edge region}
  **else if** $y_B < -z_B$ **then** case_II_3_h
        {$B$ is in the bottom edge region}
      **else** case II_3_i
        {$B$ is inside pyramid, the whole line segment is visible}
**end** {case_II_3};

### II.3.a) Point A is inside of the pyramid and point B is in top-right corner region.

The comparison of point $B$ with the plane $\rho_2$ specifies which boundary (top or right) to be used to compute the intersection point. An implementation can be as follows:

**begin** {case_II_3_a}
  **if** $((x_A+z_A)*(\Delta z-\Delta y) > (z_A-y_A)*(\Delta z+\Delta x))$
  **then** {intersection with top boundary}
    $t_1 := (y_A - z_A) / (\Delta z - \Delta y)$
  **else** {intersection with right boundary}
    $t_1 := -(x_A+z_A)/(\Delta z+\Delta x);$
  $t_2 := 0;$
**end** {case_II_3_a};

    The cases case_II_3_c, case_II_3_d and case_II_3_e are similar to case_II_3_a.

### II.3.b) Point A is inside of the pyramid and point B is in right edge region.

The appropriate intersection point (the intersection point on the right boundary) is calculated. The following pseudo-code shows how it can be implemented:

**begin** {case_II_3_b}
    $t_1 := -(x_A+z_A)/(\Delta z+\Delta x);$
    $t_2 := 0$
**end** {case_II_3_b};

    The cases case_II_3_f, case_II_3_g and case_II_3_h can be solved similarly.

### III. Point A is beyond the left boundary.

This case can be solved similarly to the case when point $A$ is beyond the right boundary (case_I).

    Finally, we can easily compute the end-points of the output line segment from the parameter value $t$ as follow:

$$x := t*\Delta x + x_A;$$
$$y := t*\Delta y + y_A;$$
$$z := t*\Delta z + z_A$$

    It can be seen that all possible cases were solved and the complete algorithm can be got by substitution all procedures by appropriate codes.

## 4. Experimental results

To be able to compare the CS, LB and CB algorithms with the proposed PC algorithm and evaluate the efficiency of the PC algorithm, we introduce three coefficients of efficiency as:

$$v_1 = \frac{T_{CS}}{T_{PC}}, v_2 = \frac{T_{LB}}{T_{PC}}, v_3 = \frac{T_{CB}}{T_{PC}}$$

where: $T_{CS}$, $T_{LB}$, $T_{CB}$ and $T_{PC}$ denote the time consumed by the CS algorithm, LB algorithm, CB algorithm and the proposed PC algorithm, respectively.

    For experimental verification, $80.10^6$ different line segments were randomly generated for each of the 21 cases shown in Figures 5 and 6. The tests were performed on the PC Intergraph Pentium-II, 400MHz, 512MB RAM, 256 KB CACHE. The obtained results are presented in Table 1, which shows that the proposed algorithm is just as fast as the CS algorithm when the line segment lies inside the pyramid, and is significantly faster in all other cases. It can be seen that the speed-up varies from 1.17 to 2.08 approximately for these cases. Table 1 also shows that the PC algorithm is significantly faster than LB and CB algorithm for the most cases.

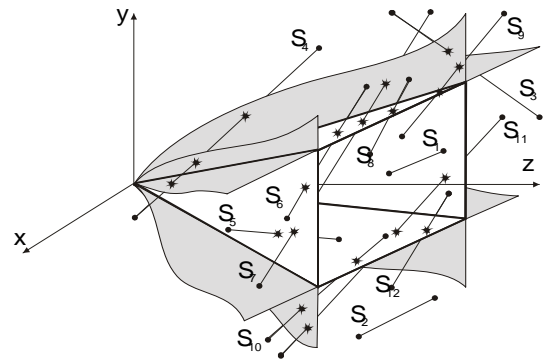

Figure 5: Generated line segments.

Figure 6: Generated line segments.

| case | $v_1$ | $v_2$ | $v_3$ |
|------|-------|-------|-------|
| s1   | **1.01** | 2.54 | 2.40 |
| s2   | **1.29** | 2.83 | 2.69 |
| s3   | **1.51** | 1.85 | 1.75 |
| s4   | **1.52** | 1.84 | 1.75 |
| s5   | **1.25** | 1.74 | 1.63 |
| s6   | **1.31** | 1.35 | 1.25 |
| s7   | **1.57** | 1.26 | 1.17 |
| s8   | **1.18** | 1.78 | 1.63 |
| s9   | **1.55** | 1.64 | 1.51 |
| s10  | **1.53** | 1.56 | 1.46 |
| s11  | **1.54** | 1.22 | 1.14 |
| s12  | **1.17** | 1.68 | 1.58 |
| s13  | **1.25** | 1.28 | 1.19 |
| s14  | **1.18** | 1.69 | 1.57 |
| s15  | **2.08** | 1.53 | 1.74 |
| s16  | **1.38** | 1.44 | 1.62 |
| s17  | **1.50** | 1.23 | 1.13 |
| s18  | **1.63** | 1.09 | 1.00 |
| s19  | **1.18** | 1.25 | 1.15 |
| s20  | **1.49** | 1.23 | 1.12 |
| s21  | **1.32** | 1.30 | 1.22 |

Table 1: Experimental results.

## 6. Conclusion

The new line segment clipping algorithm against a given pyramid in $E^3$ was developed, verified and tested. This algorithm is convenient for all applications if the line segment clipping in $E^3$ is to be used. Experiments have shown that the new algorithm is never slower than the CS, LB and CB algorithm, and is generally faster, up to twice as much in some cases.

## Acknowledgments

The authors would like to express their thanks to all who contributed to this work, especially to colleagues, recent MSc. and Ph.D. students of Computer Graphics at the University of West Bohemia in Pilsen, who stimulated this work. The special thanks belong to anonymous reviewers for very useful comments and recommendations.